# Giant Polarization Drag in a Gas of Molecular Superrotors

Uri Steinitz[1,2*], Ilya Sh. Averbukh[2*]

[1]Soreq Nuclear Research Center, Yavne, Israel

[2]AMOS and Department of Chemical and Biological Physics,
The Weizmann Institute of Science, Rehovot, Israel

* Corresponding author. E-mail: uri.steinitz@gmail.com (US); ilya.averbukh@weizmann.ac.il (IA)

September 27, 2019

## Abstract

Experiments on light dragging in a moving medium laid the cornerstones of modern physics more than a century ago, and they still are in the focus of current research. When linearly polarized light is transmitted through a rotating dielectric, the polarization plane is slightly rotated – a phenomenon first studied by Fermi in 1923. For typical non-resonant dielectric materials, the measured polarization drag angle does not surpass several microradians. Here we show that this effect may be dramatically enhanced if the light is sent to a gas of fast unidirectionally spinning molecular superrotors. Several femtosecond-laser labs have already succeeded in optically creating such a medium. We show that the specific rotation power of the superrotor medium exceeds the values previously observed in mechanically rotated bulk optical specimens by many orders of magnitude. This nonreciprocal opto-mechanical phenomenon may open new avenues for ultra-fast control of the polarization state of light.

**Main text:** Light propagating in a moving medium can be 'dragged' by it and change the speed depending on the medium motion. This was first discussed by Fresnel in 1818 when considering a hypothetical 'æther drag' [1], and studied experimentally by Fizeau [2] in a real moving dielectric medium- water flowing in the arms of an optical interferometer. In 1885 Thomson considered an electromagnetic wave traveling in a rotating æther [3], and argued that the light should experience a transverse drag leading to the rotation of the field polarization vector. While the Michelson-Morley experiment [4] essentially ruled out the notion of 'luminiferous æther', the problem of polarization drag in a real rotating matter was examined by Fermi [5], who considered it as a generalization of the Fizeau experiment. The phenomenon was then studied in depth theoretically [6, 7, 8], and was first observed experimentally by Jones [9] using a rotating glass rod. The observed polarization rotation angle (which is proportional to the angular velocity of the glass rotation, and to the time it takes the light to traverse the specimen) was very small – several microradians. Jones' experiment approached the mechanical and material strength limits, and for more than 30 years no other research managed to exceed or reproduce his result. A remarkable advance was introduced in 2011 when the rotary polarization drag was enhanced by about four orders of magnitude by using near-resonant slow-light in a rotating solid ruby rod [10]. Recently it was shown that the rotary polarization drag takes place on astronomical scale in the magnetosphere of rotating pulsars, offering a unique means to determine the rotation direction of pulsars [11]. The effect is especially important since it is nonreciprocal, i.e., creates asymmetric wave transmission between two locations.

Here we propose an alternative system for creating rotational polarization drag at unprecedentedly high specific rotation level – four orders of magnitude higher than in the above record slow-light experiment [10], and correspondingly – almost ten orders of magnitude higher than in the original Jones' observation [9]. Instead of mechanically rotating a bulky dielectric object as a whole, we suggest using short laser pulses for exciting fast unidirectional rotation of individual microscopic constituent elements of the medium. This can be hardly done for solid state samples, however can be implemented in a gas of molecular superrotors. Superrotors are molecules excited to extremely high rotational states (with $J$ of the order of tens, hundreds, and even more than a thousand of $\hbar$). Current laser techniques such as cross-polarized pulse pairs [12, 13, 14], chiral pulse trains [14], polarization shaped pulses [15], multiple kick pulses [16], and, especially, optical centrifuge [17, 18, 19, 20], can bring molecules to very fast unidirectional spinning. Gases of superrotors have been already studied both theoretically and experimentally. They have unique kinetic and optical properties [21, 22, 23], demonstrate inhibited rotational relaxation rate [24, 20, 25, 26, 21, 27, 22], persistent optical anisotropy [21, 22, 28], and giant rotational Doppler shift [29, 23, 20]. Here we show, for the first time, that such a gas is extremely optically active, with a giant specific rotation power in a broad wavelength range. This optically controllable linear nonreciprocal effect operates at standard gas conditions, and holds promise for potential applications for sensitive optical detectors and quantum information technology, among other uses.

The effect is based on the fact that the apparent frequency of circularly polarized (CP) light, as felt by a fast rotating molecule, is shifted from the frequency of the incident light due to the rotational Doppler effect [30], where the sign of this shift depends on mutual rotation directions of the optical field and the molecule. Therefore, when molecules of a gas have a biased rotation sense, dispersion results in different propagation speeds of the left and right CP beams, thus leading to a non-magnetic Faraday effect, rotating the polarization direction of passing linearly polarized light.

To illustrate the phenomenon and estimate the magnitude of the effect, we consider a gas of linear molecules uniformly distributed in space with the number density $N$. For simplicity, we assume that the molecules can rotate only in the $xy$ plane, and the initial angle $\varphi$ of the molecular axis with respect to the $x$-axis is uniformly distributed in the interval $[0, 2\pi]$. This is a reasonable approximation for molecules shortly after the end of the optical centrifugation (see a discussion below). We consider a plane CP electromagnetic wave of frequency $\omega$ and electric field amplitude $E$ propagating in the $z$-direction. The electric field polarizes a molecule rotating with angular velocity $\Omega$ and oriented at angle $\varphi + \Omega t$. We consider a simplistic model of point-like polarization charge $e$ of mass $m$ that is able to move only along the axis of the rotating linear molecules. We assume that the charge is subject to the elastic restoring force of $-m\omega_0^2 r$, where $r$ is the charge's distance from the molecular center, and $\omega_0$ is the frequency of free oscillations. In the frame rotating with the molecule, the equation of motion for the charge is

$$m\ddot{r} = eE \cos\left[(\omega + \Omega) t + \varphi\right] + m\Omega^2 r - m\omega_0^2 r \quad , \tag{1}$$

where the second term in the RHS is the centrifugal force. Here we consider the CP field rotation to be opposite to the direction of the molecular rotation, thus the electric field component along the molecular axis oscillates at the sum of the frequencies. A steady-state solution of Eq. (1) is

$$r_0(t) = \frac{eE \cos\left[(\omega + \Omega) t + \varphi\right]}{m\left(\omega_0^2 - \omega^2 - 2\Omega^2 - 2\omega\Omega\right)} \quad . \tag{2}$$

The only component of the dipole moment whose average over $\varphi$ is non-zero is the component parallel to the electric field, $d_\parallel = er_0 \cos\left[(\omega + \Omega)t + \varphi\right]$, so that the polarization density, $P$, is:

$$P = N \langle d_\parallel \rangle = \frac{Ne^2/2m}{\omega_0^2 - \omega^2 - 2\Omega^2 - 2\omega\Omega} E \quad ,$$

where the angle brackets denote averaging over $\varphi$. The relation between the displacement and the field, $D = \varepsilon_0 \varepsilon_r E = \varepsilon_0 E + P$ provides an expression for the electric susceptibility of the medium

$$\varepsilon_r^+ = 1 + \frac{Ne^2}{2\varepsilon_0 m} \frac{1}{\omega_0^2 - \omega^2 - 2\Omega^2 - 2\omega\Omega} \quad , \tag{3}$$

where the '+' superscript refers to the circularly polarized electric field counter-rotating with respect to the molecules, and $\varepsilon_0$ is the vacuum susceptibility. For the negative CP wave, whose electric field vector rotates in the same sense as the molecules, the result is similar, but with a negative $\omega$, leading to a different propagation speed. This shows that the medium exhibits circular birefringence, which affects the propagation of a linearly polarized light made of two CP components of equal amplitude, but opposite handedness. Upon propagation through a length $L$ in such a gas, the relative phase between the two CP components changes, and the light polarization vector rotates by an angle [31]:

$$\Delta\Phi = \left(\sqrt{\varepsilon_r^+} - \sqrt{\varepsilon_r^-}\right)\frac{\omega L}{2c} \quad , \quad (4)$$

(see Fig(1)). As the refractive index of the gas is only slightly different from unity, we obtain

$$\Delta\Phi \approx \frac{Ne^2 L}{2\varepsilon_0 mc}\frac{\omega^2 \Omega}{\left(\omega_0^2 - \omega^2 - 2\Omega^2\right)^2 - 4\omega^2\Omega^2} \quad . \quad (5)$$

Let us digress, and note that for non-rotating molecules ($\Omega = 0$), the phase refractive index, $n_\phi$, is (according to Eq. (3))

$$n_\phi = \sqrt{\varepsilon_r} \approx 1 + \frac{Ne^2}{4\varepsilon_0 m}\frac{1}{\omega_0^2 - \omega^2} \quad ,$$

while the group index, $n_g$ is defined by

$$n_g - n_\phi = \omega\frac{\partial n_\phi}{\partial \omega} \approx \frac{Ne^2}{2\varepsilon_0 m}\frac{\omega^2}{\left(\omega_0^2 - \omega^2\right)^2} \quad . \quad (6)$$

By keeping only linear in $\Omega \ll \omega$ term in Eq. 5, we use Eq. 6 to express the polarization drag as:

$$\Delta\Phi \approx (n_g - n_\phi)\frac{L}{c}\Omega \quad . \quad (7)$$

Equation (7) expresses the relation between the dispersion and the polarization drag, and allows assessment of the effect based on the known data regarding the refractive index of a gas of non-rotating molecules.

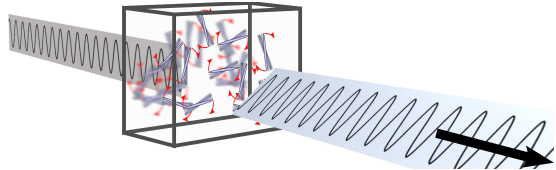

Figure 1: Linearly polarized light propagating through a gas of unidirectionally spinning molecular superrotors experiences rotation of the polarization plane.

In order to estimate the magnitude of the effect, we consider parameters typical for several published optical centrifuge setups spinning various linear molecules, like $N_2$, $O_2$, $CO_2$ or $N_2O$ [24, 20, 28, 27]. Taking oxygen superrotors [20] as an example, we note that just after the end of the centrifugation pulse, the gas demonstrates linear birefringence because of the remnant alignment of the molecules that were trapped by the centrifuge field. This angular inhomogeneity decays due to the dispersion of the rotational frequencies, and elastic collisional effects. At room temperature, the decay time of linear birefringence ranges from several tens of ps×atm (picosecond×atmosphere) for moderately high rotational excitation [32, 33, 34, 35], to hundreds of ps×atm in fully developed superrotor regime [25]. This leads to a practical loss of the linear birefringence (when probing along the rotation axis) within about a nanosecond, well before the rotational energy decays [25, 21, 22, 27]. From this moment onward, the gas is suitable for polarization rotation measurements. Consider $O_2$ molecules at ambient conditions spun to the angular velocity of $\Omega \sim 10^{14}\,rad/s$ by an optical centrifuge with an effective length of $L = 1.5\,cm$. For a probe beam of wavelength of $\sim 0.4\,\mu m$, the difference between the group and phase refractive indices is $n_g - n_\phi \sim 4\cdot 10^{-5}$ (see the Methods Section for details). This value should be multiplied by $\delta$, the fraction of the molecules that are actually captured by the centrifuge, and we estimate it as $\delta = 10\%$ (see [36] and the Methods Section). As a result, Eq. (7) estimates the polarization rotation angle as $\Delta\Phi \sim 20\,mrad \sim 1°$, which is four orders of magnitude higher than Jones' measured results. Moreover, if we consider the specific rotation $[\alpha]$, i.e., the rotatory power per density [37], we obtain $[\alpha] \sim 10^4 \deg \cdot (g/mL)^{-1} \cdot (dm)^{-1}$, a value that, when realized, would exceed the results of the slow light experiment [10] by four orders of magnitude. The immunity of the superrotors to collisions [24, 20, 28, 21, 22, 27] makes this effect stable and long lasting, allowing multiple passages of light through the superrotor area for even further increase of the observed polarization rotation angle.

To conclude, the currently available optical centrifuge technology has already succeeded in creating dense gases of molecular superrotors – molecules that rotate so fast that they are nearly immune to the collisional rotational relaxation for extended periods of time. As we showed in this paper, the superrotor gas may exhibit considerable optical activity, delivering a sizable rotation of the polarization plane of a linearly polarized probe light. An experimental test of this phenomenon can be readily implemented using already existing setups. For instance, the authors of [28] demonstrated that optical centrifuge pulses propagating in an oxygen gas cell at ambient conditions create extended narrow channels of high superrotor abundance. For several nanoseconds after the excitation pump has been switched off, the superrotors in a channel remain in the 'gyroscopic phase' [21, 22], during which the molecules behave like tiny gyroscopes, retaining the orientation of their high angular momentum even after multiple collisions. We predict that a linearly polarized probe light propagating through such a channel may twist its polarization plane by about 1°, an angle comparable with the one achieved in the experiment [10] using near resonant slow-light passing through a rotating solid ruby rod. In contrast to [10], our effect is non-resonant, and the polarization rotation affects light in a wide wavelength range.

Importantly, this is a nonreciprocal phenomenon, the same as the magnetic Faraday effect. This means that reflecting the polarized probe light back through the same superrotor medium does not undo the polarization drag: on the contrary, it doubles the effect. This opens new options for further amplification of the polarization change by multiple light passes through the gas, especially in an intra-cavity implementation. Moreover, the nonreciprocal character of the phenomenon inspires the vision of ultrafast optically switchable optical isolators utilizing molecular superrotors, and of related devices important both from the fundamental and practical points of view [38, 39]. Finally, we note that light with orbital angular momentum [40] may serve as an additional interesting probe of the polarization drag in the gas of molecular superrotors.

## Methods

In order to assess the refractive index of the gas of molecular oxygen at standard conditions (20°C, 101325 Pa), we use the formula from [41],

$$(n_\phi(\lambda) - 1) \times 10^8 = 11814.94 + \frac{970893.1}{75.4 - \lambda[\mu m]^{-2}} \quad .$$

Therefore, $n_g - n_\phi = -\lambda \partial_\lambda n_\phi \approx 2.5 \cdot 10^{-5}$ (for $\lambda = 0.4\,\mu m$). This estimate is valid for a gas with isotropic 3D angular distribution. Since the axes of the rotating molecules are confined to the $xy$ plane, their alignment factor is $\langle \cos^2 \theta \rangle = 1/2$ instead of $1/3$, the latter being the value for isotropic angular distribution ($\theta$ is the angle between the molecular axis and the field polarization direction). Thus the susceptibility of a gas of rotating superrotors is enhanced by a factor of $3/2$ compared to that of isotropic gas, as has been demonstrated in measuring the birefringence of a probe propagating perpendicular to the optical centrifuge rotation axis [42]. This brings us to assess the relevant group to phase refractive index difference as $n_g - n_\phi \sim 3.8 \cdot 10^{-5}$.

The maximum angular velocity of the oxygen superrotors may reach $10^{14}\,rad/s$ (based on the observed rotational quantum number of 130 [43], or from the rotational Raman spectra of centrifuged $O_2$ molecules shown in [42]). To evaluate the fraction of all molecules trapped by the centrifuge, we estimate the depth of the centrifuge potential well as $U \sim \Delta\alpha E_c^2/4$, where $\Delta\alpha = 1.12 \text{Å}^3$ is the polarizability anisotropy of the oxygen molecules, and $E_c$ is the amplitude of the optical electric field of the centrifuge pump. A centrifuge pulse of the intensity of $10^{12}\,W/cm^2$ to $5 \cdot 10^{12}\,W/cm^2$ affects up to three lowest rotational levels of the oxygen molecules. The total relative thermal population of the affected states ranges from 4% to 26% at 300K, depending on the centrifuge intensity. We assume that only 10% of the molecules are captured by the optical centrifuge potential well, the rest rotate randomly. This estimate is consistent with the one provided in [36].

The Rayleigh length of the centrifuge field is estimated using the reported value of the radius of the pump beam's profile at the focal plane, $w_0 = 45\,\mu m$ [43]. Subsequently, the effective length $L$ is twice the Rayleigh length, $L \sim 2 \cdot \pi w_0^2/\lambda \sim 1.5\,cm$ for $\lambda \sim 0.8\,\mu m$.

The expected specific rotation $[\alpha]$ for a gas of oxygen superrotors described in this paper, is calculated by dividing the angle of polarization rotation ($\sim 1°$) by the active length ($\sim 1.5\,cm$) and by the mass density of the medium ($\sim 1.4\,g/L$). Using standard units for $[\alpha]$ [37], we obtain $[\alpha] \sim 10^4\,\deg \cdot (g/mL)^{-1} \cdot (dm)^{-1}$. For comparison, in Jones' experiment [9] (where the rotation of $\sim 3\,\mu rad$ was obtained in about $40\,cm$ of glass of supposedly $\sim 5\,g/mL$) $[\alpha] \sim 10^{-6}\,\deg \cdot (g/mL)^{-1} \cdot (dm)^{-1}$. Franke-Arnold *et al* [10] reported rotation of several degrees in about $10\,cm$ of ruby, reaching a specific rotation value in the order of $[\alpha] \sim 1\,\deg \cdot (g/mL)^{-1} \cdot (dm)^{-1}$.

# Acknowledgments

The authors are grateful to V. Milner and A.A. Milner for many useful discussions. Funding: This work was supported by the Israel Science Foundation (Grant No. 746/15). I.A. acknowledges support as the Patricia Elman Bildner Professorial Chair. This research was made possible in part by the historic generosity of the Harold Perlman Family. Authors contributions: The authors contributed equally to this work. Competing interests: The authors have no competing interests. Data and materials availability: All data is available in the manuscript or the supplementary materials.

# References

1. A. Fresnel, *Annals of chemistry and physics* **9**, 57 (1818).

2. A.-H. Fizeau, *Comptes Rendus de l'Acadèmie des Sciences* **33**, 349 (1851).

3. J. J. Thomson, *Proc. Camb. Phil. Soc.* **5**, 250 (1885).

4. A. A. Michelson, E. W. Morley, *Am. J. Sci.* **34**, 333 (1887).

5. E. Fermi, *Rend. Lincei* **32**, 115 (1923).

6. M. Player, *Proc. Royal Soc. Lond. A* **349**, 441 (1976).

7. M. W. Evans, *International Journal of Modern Physics B* **6**, 3043 (1992).

8. G. Nienhuis, J. Woerdman, I. Kuščer, *Phys. Rev. A* **46**, 7079 (1992).

9. R. V. Jones, *Proc. Royal Soc. Lond. A* **349**, 423 (1976).

10. S. Franke-Arnold, G. Gibson, R. W. Boyd, M. J. Padgett, *Science* **333**, 65 (2011).

11. R. Gueroult, Y. Shi, J.-M. Rax, N. J. Fisch, *Nat. Commun.* **10**, 3232 (2019).

12. S. Fleischer, Y. Khodorkovsky, Y. Prior, I. Sh. Averbukh, *New J. Phys.* **11**, 105039 (2009).

13. K. Kitano, H. Hasegawa, Y. Ohshima, *Phys. Rev. Lett.* **103**, 223002 (2009).

14. S. Zhdanovich, *et al.*, *Phys. Rev. Lett.* **107**, 243004 (2011).

15. G. Karras, *et al.*, *Phys. Rev. Lett.* **114**, 103001 (2015).

16. J. P. Cryan, P. H. Bucksbaum, R. N. Coffee, *Phys. Rev. A* **80**, 063412 (2009).

17. J. Karczmarek, J. Wright, P. Corkum, M. Ivanov, *Phys. Rev. Lett.* **82**, 3420 (1999).

18. D. M. Villeneuve, *et al.*, *Phys. Rev. Lett.* **85**, 542 (2000).

19. L. Yuan, S. W. Teitelbaum, A. Robinson, A. S. Mullin, *Proc. Natl. Acad. Sci. U. S. A.* **108**, 6872 (2011).

20. A. Korobenko, A. A. Milner, V. Milner, *Phys. Rev. Lett* **112**, 113004 (2014).

21. Y. Khodorkovsky, U. Steinitz, J.-M. Hartmann, I. Sh. Averbukh, *Nat. Commun.* **6**, 7791 (2015).

22. U. Steinitz, Y. Khodorkovsky, J.-M. Hartmann, I. S. Averbukh, *ChemPhysChem* **17**, 3795 (2016).

23. U. Steinitz, Y. Prior, I. Sh. Averbukh, *Phys. Rev. Lett.* **112**, 013004 (2014).

24. C. Toro, Q. Liu, G. O. Echebiri, A. S. Mullin, *Mol. Phys.* **111**, 1892 (2013).

25. A. A. Milner, A. Korobenko, J. W. Hepburn, V. Milner, *Phys. Rev. Lett.* **113**, 043005 (2014).

26. M. J. Murray, *et al.*, *J. Phys. Chem. A* **119**, 12471 (2015).

27. M. J. Murray, H. M. Ogden, C. Toro, Q. Liu, A. S. Mullin, *ChemPhysChem* **17**, 3692 (2016).

28. A. A. Milner, A. Korobenko, K. Rezaiezadeh, V. Milner, *Phys. Rev. X* **5**, 031041 (2015).

29. O. Korech, U. Steinitz, R. J. Gordon, I. Sh. Averbukh, Y. Prior, *Nat. Photonics* **7**, 711 (2013).

30. B. A. Garetz, *J. Opt. Soc. Am.* **71**, 609 (1981).

31. G. R. Fowles, *Introduction to modern optics* (Dover Publications (New York), 1989).

32. T. Vieillard, F. Chaussard, D. Sugny, B. Lavorel, O. Faucher, *J. Raman Spectrosc.* **39**, 694 (2008).

33. N. Owschimikow, *et al.*, *J. Chem. Phys.* **133**, 044311 (2010).

34. T. Vieillard, *et al.*, *Phys. Rev. A* **87**, 023409 (2013).

35. R. Damari, D. Rosenberg, S. Fleischer, *Phys. Rev. Lett.* **119**, 033002 (2017).

36. J. Floß, C. Boulet, J.-M. Hartmann, A. A. Milner, V. Milner, *Phys. Rev. A* **98**, 043401 (2018).

37. J. Rumble, *CRC Handbook of Chemistry and Physics, 99th Edition* (Boca Raton: CRC Press, 2018).

38. R. J. Potton, *Rep. Prog. Phys.* **67**, 717 (2004).

39. R. Fleury, D. L. Sounas, C. F. Sieck, M. R. Haberman, A. Alù, *Science* **343**, 516 (2014).

40. L. Allen, S. M. Barnett, M. J. Padgett, *Optical angular momentum* (Bristol, UK: Institute of Physics Publishing, 2003).

41. P. Křen, *Appl. Opt.* **50**, 6484 (2011).

42. A. Korobenko, *J. Phys. B: At., Mol. Opt. Phys.* **51**, 203001 (2018).

43. A. Milner, A. Korobenko, V. Milner, *Opt. Express* **23**, 8603 (2015).